\documentstyle[aps,prl,multicol,epsfig]{revtex}
\begin{document}
\title{\bf Supercooled confined water and the Mode Coupling 
crossover temperature}
\author{ P.~Gallo${}^\dagger$, M.~Rovere${}^\dagger$, 
E.~Spohr${}^\ddagger$ \\}
\address{${}^\dagger$ Dipartimento di Fisica ``E. Amaldi'', 
Universit{\`a} ``Roma Tre'', \\ 
and Istituto Nazionale per la Fisica della Materia,  Unit{\`a} di Ricerca
Roma Tre, \\ Via della Vasca Navale 84, I-00146 Roma, Italy \\
${}^\ddagger$ Department of Theoretical
Chemistry, University of Ulm, \\ Albert-Einstein-Allee 11, D-89069 Ulm,
Germany }

\date{\today}
\maketitle

\begin{abstract} 

  We present a Molecular Dynamics study of the single
  particle dynamics of supercooled water confined in a silica pore. 
  Two dynamical regimes are found: close to the hydrophilic 
  substrate molecules are below the Mode Coupling
  crossover temperature, $T_C$, already at ambient temperature. 
  The water closer to the center of the pore (free water) 
  approaches upon supercooling $T_C$ as predicted by Mode Coupling Theories.  
  For free water the crossover temperature 
  and crossover exponent $\gamma$  are extracted from 
  power-law fits to both the diffusion coefficient 
  and the relaxation time of the late $\alpha$ region.

\end{abstract}

\begin{multicols}{2}

  The effect of supercooling on the dynamics of liquids in confined
  environments is a research field that has become more and more
  popular over the last few years. Out of an extremely rich 
  phenomenology that shows diversification of specific behavior
  depending on the size of the particles, the confining geometry and
  the specific interaction with the substrate, some general trends can
  none the less be extracted~\cite{mckenna,kremer,melni}.  In fact two
  competing effects seem to be the main contributions to the
  modification of the dynamics of the confined liquid with respect to
  the bulk phase: the bare geometric confinement and the interaction
  with the substrate.  In particular there is evidence from
  experiments that liquid-wall interactions can lead to a layering
  and a decrease of mobility close to the substrate, with a substantial
  increase of the glass transition temperature. This effect is
  stronger for attractive interactions between the substrate and the
  liquid.  In some liquids experimental evidences show two distinct
  dynamical regimes~\cite{mckenna,melni}.

  Among liquids water plays a most fundamental role on earth.  The
  study of the dynamics of water at interfaces or confined in
  nanopores as a function of temperature and hydration level is
  relevant in understanding important effects in systems of interest
  to biology, chemistry and geophysics~\cite{lynden-bell,doster}.

  In particular the single particle dynamics of water confined in
  nanopores have been studied by different experimental techniques
  such as neutron diffraction and nuclear magnetic
  resonance~\cite{zanotti}.  A slowing down of the dynamics with
  respect to the bulk phase is observed.  Nevertheless the details of
  the microscopic dynamic behavior of confined water  upon supercooling
  are still unclear.

  Below $235$ K bulk liquid water enters the so called {\em no
  man's land}~\cite{stanley}, where nucleation processes, most
  likely triggered by the presence of impurities, take place and drive the
  liquid to the solid crystalline phase, preventing the experimental
  approach to the glass phase~\cite{angell,pablo}.  
  The dynamical behavior of the bulk water simulated
  upon supercooling with the use of the Simple Point
  Charge/Extended (SPC/E) site model potential~\cite{spce},
  fits in the framework of
  the idealized version of Mode Coupling Theory (MCT)~\cite{goetze}, 
  predicting a temperature of ideal structural arrest, or crossover 
  temperature $T_C$, that coincides with the so called
  singular temperature of water~\cite{speedyangell}.
  This behaviour has been observed several years ago by computer
  simulation~\cite{gallo-prl} and 
  substantiated by experimental signatures~\cite{exp-superc} and
  further simulation and theoretical works~\cite{Starr,Linda}. 
  When a liquid approaches the crossover temperature $T_C$
  MCT predicts that the dynamics is dominated by the cage
  effect.  After an initial ballistic motion, the particle is trapped
  in the transient cage formed by its nearest neighbours.  Once the
  cage relaxes the particle enters the Brownian diffusive regime.
  Below $T_C$, according to the idealized version of MCT,
  the system becomes non-ergodic. In real structural glasses
  hopping processes restore ergodicity. $T_C$ is therefore a 
  crossover temperature from a liquid-like to a solid-like regime. 
  These activated processes are not relevant above $T_C$ for most liquids.

  Until now, no systematic computer simulation studies of the 
  microscopic dynamics of confined water upon supercooling have been attempted.

  Thermometric studies~\cite{takamuku,ewhansen1}, NMR
  spectroscopy~\cite{takamuku,ewhansen1,ewhansen2,stapf}, neutron
  diffraction~\cite{takamuku,mcbf1,dore1} and X-ray
  diffraction~\cite{morishige} show evidence that two types of water
  are present in the confining pores, {\it free water} which is in the
  middle of the pore and {\it bound water} which resides close to the
  surface. Free water is observed to freeze abruptly in the cubic ice
  structure.  Bound water freezes gradually but it does not
  make any transition to an ice phase~\cite{morishige}.  Layering
  effects of water close to the substrates have been observed in all
  the simulations for different geometry and water-substrate
  interaction~\cite{lynden-bell,geiger}.

  In this letter we present evidence from Molecular Dynamics (MD)
  simulations that water confined in a hydrophilic nanopore exhibits
  two distinct dynamical regimes. In particular the fraction of free
  water molecules behaves similarly to the bulk phase, and its
  dynamics is consistent with several MCT predictions.
  Power laws fits based on MCT yield the crossover
  temperature of the fraction of free water molecules;
  simultaneously, we show that the bound water molecules are already
  below $T_C$ at room temperature.
          
  In our simulations water has been confined in a silica cavity
  modeled to represent the average properties of the pores of Vycor
  glass~\cite{jmliq1}.  Water-in-Vycor is a system of particular
  interest~\cite{mar1}, since the porous silica glass is characterized
  by a quite narrow pore size distribution with an average diameter of
  $40$ \AA{}. The pore size does not depend on the hydration level and
  the surface of the pore is strongly hydrophilic. Moreover the
  water-in-Vycor system can be considered as a prototype representing
  more complex environments of interfacial water.  Different from
  previous work on the dynamics of water close to a planar regular
  silica surface~\cite{rossky2} we use a cylindrical geometry with a
  corrugated surface.  We have constructed a cubic cell of silica
  glass by the usual quenching procedure.  As described in detail in
  a previous work~\cite{jmliq1} inside the cube of length $L=71.29$
  \AA{} we cut out a cylindrical cavity of diameter $40$ \AA{} 
  and height $L$ by eliminating all the atoms lying 
  within a distance $R = 20$ \AA{} from
  the axis of the cylinder, which is taken as the $z$-axis. After
  elimination of silicon atoms with less than four oxygen neighbours,
  we saturate the dangling bonds of oxygen atoms with hydrogen atoms,
  in analogy to the experimental situation~\cite{mar1}.

  Water molecules described by the SPC/E model are introduced into the
  cavity.  The water sites interact with the atoms of the rigid matrix
  by means of an empirical potential model, where different fractional
  charges are assigned to the atomic sites of the silica glass and
  where the oxygen sites of water additionally interact with the
  oxygen atoms of the substrate via Lennard-Jones
  potentials~\cite{jmliq1,brodka}. The MD calculations are performed
  with periodic boundary conditions
  along the $z$-direction and the temperature is controlled via coupling
  to a Berendsen thermostat~\cite{berendsen}; the shifted force method
  is used with a cutoff at $9$ \AA{} to truncate long-range
  interactions~\cite{cutoff}.

  In this work we consider water at a density corresponding to the
  experimentally determined level of full hydration~\cite{mar1}.
  In the chosen geometry this corresponds to $N_w=2600$ water
  molecules and to a density $\rho=0.867$ $g/cm^3$.  
  We investigate the dynamical behavior of the confined
  water for five temperatures, namely $T = 298, 270, 240, 220$ and
  $210$ K.
%%%%%%%%%%%%%%%%%%%%%%%%%%%%%%%%%%%%%%%%%%%%%%%%%%%%%%%%%%%%%%
\begin{figure}[t]
%\setlength{\epsfxsize}{100mm}\epsfbox[75 260 380 410]{fig1.eps}
%\vspace{8.truecm}
\centering\epsfig{file=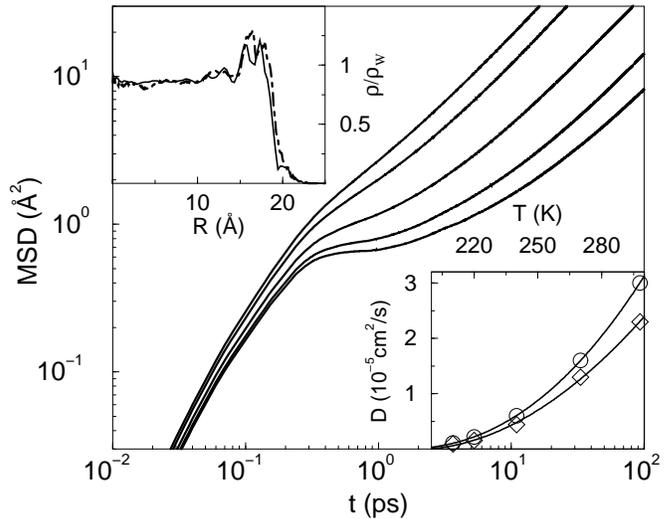,width=1\linewidth}
\caption{
  Mean square displacement (MSD) along the $z$-direction of free water
  for temperatures (from top) $T=298,270, 240,
  220,210$ K at full hydration ($N_w=2600$); in the upper
  left inset the density profiles at $T=298$ K (continuous line)
  and $T=210$ K (dot-dashed line) along the pore
  radius are shown; the lower right inset shows the diffusion
  coefficient $D$ vs temperature $T$ for the direction $z$ (open
  circles) and $xy$ (open diamonds).  Full lines are power law fits to
  the data, given by $D=9.03(T/185.3-1)^{2.21}$ and
  $D=7.70(T/194.6-1)^{1.90}$ for the $z$ and $xy$ directions,
  respectively. $D$ is in $\rm{cm^2/s}$ and $T$ is in K.  }
\protect\label{fig:1}
\end{figure} 
%__________________________________________________________________________
%%%%%%%%%%%%%%%%%%%%%%%%%%%%%%%%%%%%%%%%%%%%%%%%%%%%%%%%%%%%%%%%%%%%%%%
  In the following we focus on the single particle dynamics of the
  water molecules contained in the pore and test some of the main
  predictions of MCT. 
  The radial density profile of the water oxygen atoms, normalized to
  bulk water density, is displayed for $T=298$ K and
  $T=210$ K in the inset at the upper
  left corner of Fig.~1. It shows the high hydrophilicity of the pore
  in the form of density oscillations close to the substrate.
  These oscillations are not sensitive to supercooling.
  The average density of water molecules at $T=298$ K for 
  $0<R<15$~\AA{} is $\rho=0.897$ $g/cm^3$, 
  for $15<R<18$~\AA{} is $\rho=1.079$ $g/cm^3$ 
  and for $R>18$~\AA{} (a depletion layer) is $\rho= 0.493$ $g/cm^3$,
  where $R=\sqrt{(x^2+y^2)}$.
  In a preliminary analysis done at ambient temperature as a function
  of hydration level~\cite{euro} we found that, 
  due to the presence of strong inhomogeneities in our 
  system, a fit of the total correlators to an analytic shape can be
  carried out only by excluding the subset of molecules 
  in the double layer close to the substrate ($R > 15$~\AA{})
  which we now identify with the so called bound water. 
  In the following we show that this shell analysis keeps its validity
  upon supercooling.
  We analyze the mean square
  displacement (MSD) and the self intermediate scattering function
  (ISF) separately for bound water
  and for the remaining inner layers ($R > 15$~\AA{}),
  which we identify with the free water. 
  Since no asymptotic free motion is possible in
  the $xy$ plane, we separately analyze the dynamics within this plane
  and along the pore $z$-axis.  
  In the main frame of Fig.~1 the MSD of free water is
  displayed along the non-confined $z$-direction for the investigated
  temperatures. As supercooling progresses a plateau region due to the
  cage effect develops after the initial ballistic motion, starting
  around $t=0.3$ ps.  The MSD in the $xy$-direction (not shown) is
  characterized by a similar trend, but the dynamics is slower. In the
  low right corner the inset shows the behavior of the diffusion
  coefficient $D$ extracted from the fit of the MSD in the Brownian 
  regime region for both the $z$ and the $xy$ direction together
  with fits of the MCT predicted power law $D\sim(T-T_C)^{\gamma}$ to
  the $5$ data points. In the $z$-direction, we obtain $T_C\simeq
  185.3$ K and $\gamma\simeq 2.21$, which are similar to the values
  found for the SPC/E bulk water for ambient pressure,
  namely $T_C\simeq 186.3$ K and $\gamma\simeq 2.29$~\cite{gallo-prl,nota}.
  In the $xy$-direction, where the
  dynamics is slower, the critical temperature of free
  water increases to $T_C\simeq 194.5$ K and $\gamma\simeq 1.90$.
%%%%%%%%%%%%%%%%%%%%%%%%%%%%%%%%%%%%%%%%%%%%%%%%%%%%%%%%%%%%%%
\begin{figure}[t]
\centering\epsfig{file=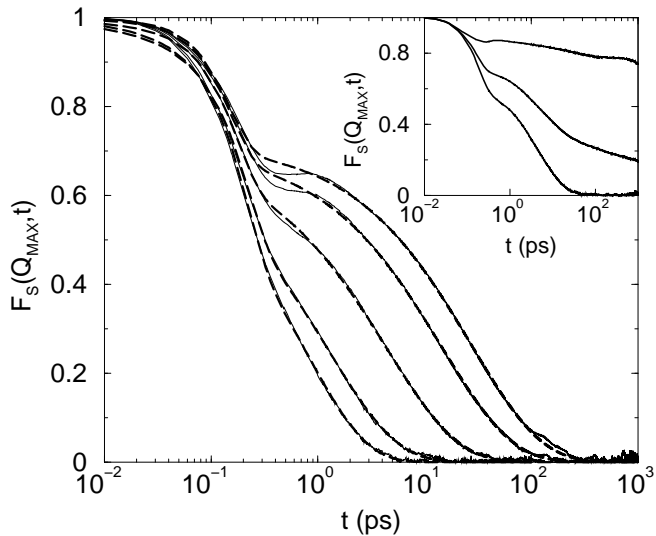,width=1\linewidth}
%\setlength{\epsfxsize}{100mm}\epsfbox[75 260 380 410]{fig2.eps}
%\vspace{8.truecm}
\caption{
  Intermediate scattering function (ISF) for free water in the
  $xy$-direction at the peak of the structure factor ($Q_{MAX} =
  2.25$~\AA${}^{-1}$)
  for the five investigated temperatures. 
 Curves on the top correspond to lower temperatures.
 Full lines are the MD
  data and long-dashed lines are the fit by
  Eq.~(\protect\ref{strexp}). In the inset the full layer analysis is
  shown for $T=240$ K. The central curve is the total ISF, the upper
  curve is the bound water contribution and the lower curve is the
  free water contribution to the total ISF.  } \protect\label{fig:2}
\end{figure} 
%%%%%%%%%%%%%%%%%%%%%%%%%%%%%%%%%%%%%%%%%%%%%%%%%%%%%%%%%%%%%%%%%%%%%%%
  In Fig.~2 we show the ISF of free water at the peak of the
  oxygen-oxygen structure factor along the $xy$-direction,
  $Q_{MAX}=2.25$ $\rm\AA{}^{-1}$, as a function of temperature. The free
  water molecules inside the pore show, similar as in SPC/E bulk water, a
  diversification of the relaxation times as supercooling proceeds.
  The plateau region stretches as $T_C$ is approached.  The long time
  region, the so-called late $\alpha$ region, is expected to have a
  stretched exponential decay for a liquid approaching $T_C$. 
  In the same figure, the fit of the function
\begin{equation}
  F_S(Q,t)= \left[ 1-A(Q) \right] e^{-\left( t/\tau_s \right)^2}+
  A(Q)e^{- \left( t/\tau_l \right)^\beta}
\label{strexp}
\end{equation}
to the data points is shown, where $A(Q)=e^{-a^2Q^2/3}$ is the
Debye-Waller factor arising from the cage effect with $a$ the
effective cage radius. $\tau_s$ and $\tau_l$ are,
respectively, the short and the long relaxation times, and $\beta$ is
the Kohlrausch exponent.  Obviously the Gaussian form
of the fast relaxation can be only an approximate one.

In the inset of Fig.~2 we show the full layer analysis for $T=240$ K
as a representative case. The topmost curve shows the behavior of
bound water, while the lower curve is the ISF of free water (identical
to the curve in the main picture). It is clearly seen that bound water
is below $T_C$, since the correlation function does
not decay to zero on the nanosecond time scale.  The central curve is
the total ISF of confined water. It displays a strong
non-exponential tail, which cannot be fitted by a stretched
exponential function. Our layer analysis shows clearly
that the contribution of free water can be separated from the one of
bound water and that the stretched exponential function is able to
give a very good fit to the late part of the $\alpha$ region in the
free water subsystem as we supercool. 

%%%%%%%%%%%%%%%%%%%%%%%%%%%%%%%%%%%%%%%%%%%%%%%%%%%%%%%%%%%%%%%%%%%%%%%%%%%%
\begin{figure}[t]
\centering\epsfig{file=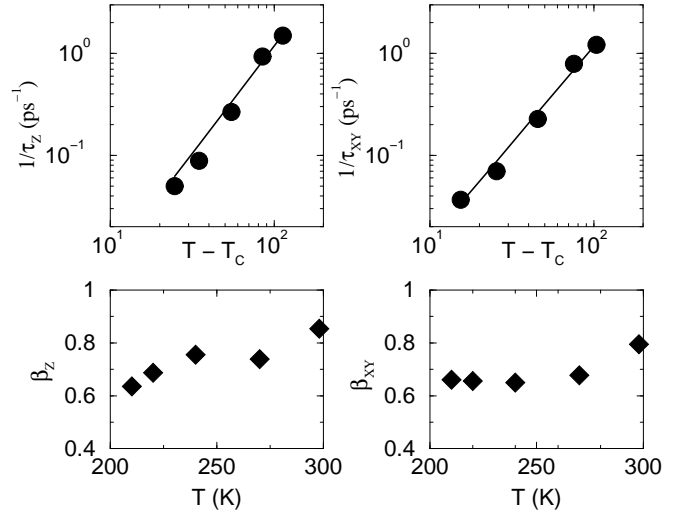,width=1\linewidth}
%\setlength{\epsfxsize}{100mm}\epsfbox[75 260 380 410]{fig3.eps}
%\vspace{8.truecm}
\caption{
  The upper part of the figure shows log-log plots of
  the inverse relaxation times, $\tau_l^{-1}$, as a function of $(T-T_C)$
  along the $z$-direction (left) and the $xy$ direction (right).
  The full lines are power law fits given by ${1}/{\tau_z} \sim
  (T-185.3)^{2.11}$ on the left and ${1}/{\tau_{xy}} \sim
  (T-194.6)^{1.90}$ on the right. The lower part shows the $T$
  dependence of the Kohlrausch exponents $\beta$ along the $z$ (left)
  and $xy$ directions (right).  } \protect\label{fig:3}
\end{figure} 
%__________________________________________________________________________
%%%%%%%%%%%%%%%%%%%%%%%%%%%%%%%%%%%%%%%%%%%%%%%%%%%%%%%%%%%%%%%%%%%%%%%

The values of the Kohlrausch exponents $\beta$ and the relaxation
times $\tau_l$ extracted from the fits of Eq.~(\ref{strexp}) to the
ISF data (Fig.~2) are reported in Fig.~3 for free water as a function
of temperature for the $z$ and the $xy$ directions. 
The $T$ dependence of $\beta$ and $\tau_l$ is in agreement with MCT.
MCT also predicts that
the inverse of the $\alpha$-relaxation time $\tau_l$ vanishes with
the same power law as the diffusion coefficient
$1/\tau_l\sim(T-T_C)^\gamma$.  In the upper part of Fig.~3 we show
log-log plots of the inverse relaxation times as function of $(T-T_C)$
as points together with power law fits as continuous lines.  From the
fits we obtain values very close to the ones obtained above from the 
power law fit  of the diffusion coefficients (see inset in Fig.~1).
This result seems therefore in agreement with 
the MCT prediction that the $\gamma$
exponent should be independent of the quantity investigated.
In pure SPC/E water the quantity $D\tau$ was found relatively
constant along isochores. None-the-less a slight increase of the product 
was found on cooling most likely due to the progressive breakdown of MCT
on approaching $T_C$, see Fig. 5c of Ref.\cite{Starr}. 
In our case a definitive conclusion on the coupling of $D$ and $\tau$
cannot be made at the present stage due to the fluctuations in the data.
The cage radii extracted from the value of the Debye-Waller factor
along both the $xy$ and $z$ directions range from $a=0.51$ \AA{} for
$T=298$ K to $a=0.45$ \AA{} for $T=210$ K.  The $\tau_s$ values 
are $\tau_s \sim 0.2$~ps. 
These values are very similar to those of SPC/E bulk 
water~\cite{gallo-prl}.

In summary, we have presented evidence that the dynamical behavior of
SPC/E water confined in a silica nanopore upon supercooling
can be analyzed in terms of two subsets of water
molecules with clearly distinct dynamical regimes,
in agreement with signatures found in experimental 
studies on confined liquids~\cite{mckenna,kremer,melni}.
Due to the presence of a strongly attractive surface we do not find,
at variance with MD studies on other confined liquids, 
density oscillations all over the confining space~\cite{lowen},
or a continuous behaviour in going
from the center to the substrate~\cite{scheidler}.
The bound water being close to a strong hydrophilic surface suffers a severe
slowing-down already at room temperature. 
Power laws fit based on MCT predict that for free water ideal structural arrest
would occur at $T_C$ $186.3K$ and $194.5$ respectively along $z$ and $xy$ 
directions for the full hydration level of the pore. 
For the investigated quantities dynamics appears well accounted for by the
idealized MCT of supercooled liquids.  In this respect the predictions
of the idealized MCT appear to be robust and able to describe also
confined molecular liquids provided that the effects of the
interaction with the substrate are properly taken into account.
Experimental evidence of a possible MCT behavior of water-in-Vycor
have been observed~\cite{zanotti,chen1}.
Therefore the analysis presented here represents an important step
towards the understanding of slow structural relaxation of highly non
trivial glass forming systems.

M.~R. and P.~G. acknowledge the financial support of the
G Section of the INFM.

\end{multicols}

\begin{references}
\bibitem{cutoff}
We did check that the use of a larger cutoff or Ewald
summations does not change the trend of the obtained results, 
see discussion in ref.~\onlinecite{jmliq1}.
\end{references}
\end{document}